\documentclass[a4paper,onecolumn,superscriptaddress,11pt,accepted=2021-07-05]{quantumarticle}
\usepackage[utf8]{inputenc}
\usepackage[english]{babel}
\usepackage[numbers,sort&compress]{natbib}
\usepackage{hyperref}
\usepackage{xcolor}
\usepackage{graphicx}
\usepackage{amsfonts,amsmath,amssymb,amsthm}
\usepackage{tikz}
\usepackage{pgfplots}
\usepackage{authblk}
\usepackage[title]{appendix}
\usepackage{bbold}

\newcommand{\proj}[1]{\ketbra{#1}{#1}}
\newcommand{\ket}[1]{\left|#1\right\rangle}
\newcommand{\bra}[1]{\langle#1|}

\newcommand{\ketbra}[2]{|#1\rangle\langle#2|}

\newcommand{\smb}{\smallbreak}  % skip a small line
\newcommand{\mb}{\medbreak}  % skip a line

\DeclareMathOperator{\Tr}{Tr}

\title{Single-copy activation of Bell nonlocality via broadcasting of quantum states}

\author[1]{Joseph Bowles}
\author[2]{Flavien Hirsch}
\author[1]{Daniel Cavalcanti}
\affil[1]{ICFO-Institut de Ciencies Fotoniques,  The Barcelona Institute of Science and Technology,  08860 Castelldefels (Barcelona),  Spain}
\affil[2]{Institute for Quantum Optics and Quantum Information (IQOQI), Austrian Academy of Sciences, Boltzmanngasse 3, 1090 Vienna,
Austria}

\date{\today}

\begin{document}

\maketitle

\begin{abstract}
    Activation of Bell nonlocality refers to the phenomenon that some entangled mixed states that admit a local hidden variable model in the standard Bell scenario nevertheless reveal their nonlocal nature in more exotic measurement scenarios. We present such a scenario that involves broadcasting the local subsystems of a single-copy of a bipartite quantum state to multiple parties, and use the scenario to study the nonlocal properties of the two-qubit isotropic state: 
    \begin{align}
    \nonumber \rho_\alpha = \alpha\,\proj{\Phi^+}+(1-\alpha)\frac{\mathbb{1}}{4}. 
    \end{align}
    We present two main results, considering that Nature allows for (i) the most general no-signalling correlations, and (ii) the most general quantum correlations at the level of any hidden variable theory. We show that the state does not admit a local hidden variable description for $\alpha>0.559$ and $\alpha>\frac{1}{2}$, in cases (i) and (ii) respectively, which in both cases provides a device-independent certification of the entanglement of the state. These bounds are significantly lower than the previously best-known bound of $0.697$ for both Bell nonlocality and device-independent entanglement certification using a single copy of the state. Our results show that strong examples of non-classicality are possible with a small number of resources.
\end{abstract}

\tableofcontents

\section{Overview}
The discovery by John Bell that quantum theory cannot be reproduced by local hidden variables (LHVs) is one of the most fascinating outcomes of modern physics and is made possible by the strange nature of quantum correlations. More precisely, measurement outcomes arising from local measurements made on entangled quantum systems exhibit correlations which provably forbid any explanation in terms of LHVs \cite{Bell64,BrunnerReview}. This phenomenon---termed \emph{Bell nonlocality}---can be experimentally witnessed via the violation of Bell inequalities: functions of the measurement statistics that are bounded for LHV theories. Since Bell's original theorem, many experiments have observed Bell inequality violations,  proving that Nature---quantum or otherwise---evades any LHV description \cite{BrunnerReview,Freedman,Aspect,Hensen,bigbelltest,quasar}.

\begin{figure}
    \centering
    \includegraphics[scale=0.9]{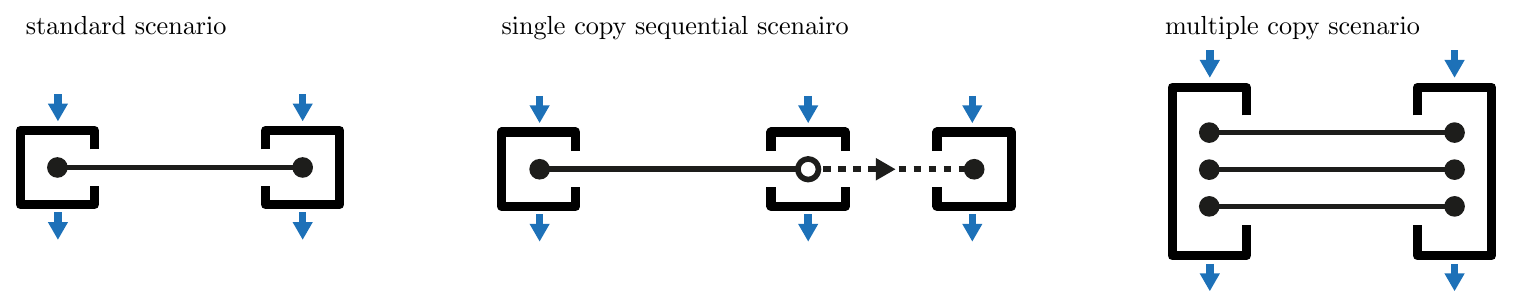}
    \caption{Scenarios for showing nonlocality of quantum states. Arrows in and out of boxes denote measurement inputs and outputs. Left: the standard Bell scenario.  centre: The single-copy sequential scenario used to show `hidden nonlocality'. Here, at least one of the local subsystems undergoes a time-ordered sequence of measurements. Right: the multiple-copy scenario. Many copies of the state are prepared and joint measurements are made between the local subsystems.}
    \label{fig:scenarios}
\end{figure}

\smb
In order to observe Bell nonlocality one performs a Bell test. Here, a source produces a bipartite quantum state that is sent to two separated devices, and a number of possible measurements are performed locally on the state. To show Bell nonlocality one violates a Bell inequality, which proves that the resulting statistics cannot be reproduced by a source of classical randomness, i.e.\ by LHVs. This scenario, that we call the \emph{standard scenario}, is shown in figure \ref{fig:scenarios}, left. %At the heart of Bell nonlocality is a proof that this quantum state cannot be replaced by a source of local hidden variables, and is witnessed by the violation of a Bell inequality. 
In the standard scenario, all pure entangled states violate a Bell inequality \cite{gisin91}, however for mixed states the situation is more subtle. Notably, there exist a large class of mixed entangled states---called Bell-local states---that do not violate any Bell inequality, due to the explicit construction of LHV models that reproduce the outcome statistics of all possible local measurements made on them \cite{Werner,Barrett02,lhvreview,bowles_sufficient,hirsch2016algorithmic,cavalcanti2016general,jevtic2015einstein}.

\smb

\begin{figure}
    \centering
    \includegraphics[scale=0.9]{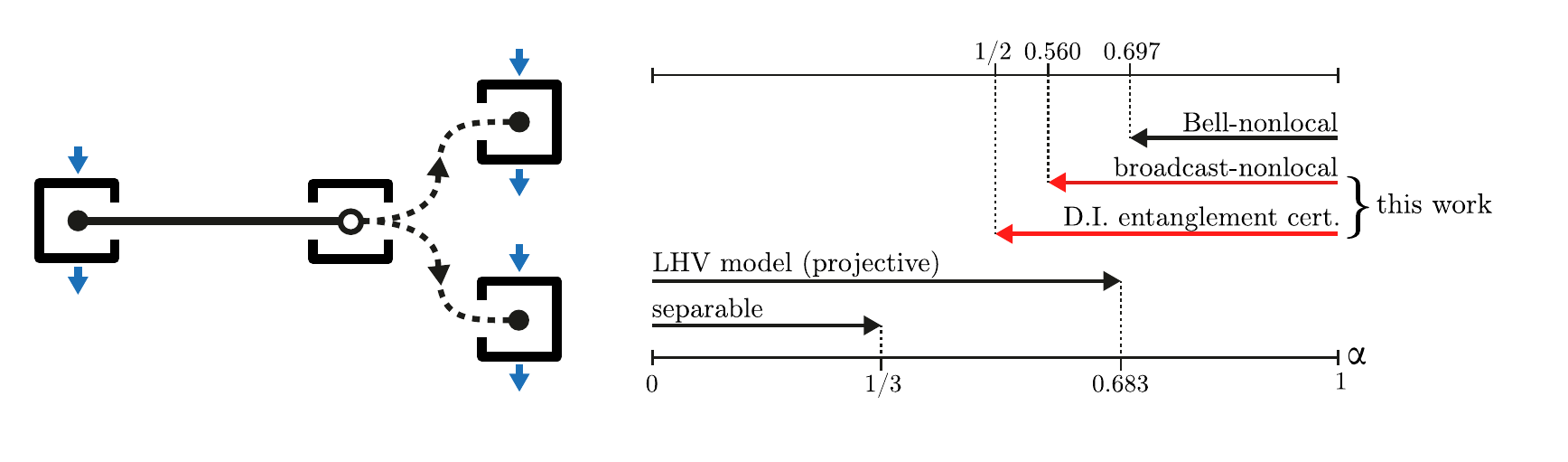}
    \caption{Left: the broadcasting scenario. One (or more) of the local systems is broadcast via the application of a quantum channel, resulting in a multipartite state. Local measurements are then performed on this state, and the resulting statistics are used to rule out a local hidden variable description for the original bipartite state. Right: 
    Bounds for for the two-qubit isotropic state \eqref{isotropic}. In this work we give bounds for which the state does not have a local hidden variable description in any no-signalling hidden variable theory (broadcast-nonlocal) and for which the entanglement of the state can be certified device-independently (D.I.\ entanglement cert.).}
    \label{fig:main}
\end{figure}

For some time it was thus presumed that Bell-local states could not exhibit Bell nonlocal behaviour. This intuition has since been proven wrong due to the phenomenon that is generally known as the \emph{activation of Bell nonlocality}  \cite{Palazuelos12,PopescuHNL,DaniSA,activation_zukowski,DaniNetwork,FlaHNL,cavalcanti2012nonlocality,bowles2016genuinely,Gallego14}. Here, one subjects the quantum state to a more complex measurement scenario, which is able to prove a lack of LHV description for some (although still not all) states  that are local in the standard scenario. Such scenarios typically come in two flavours (see figure \ref{fig:scenarios}):
\begin{enumerate}
    \item \textbf{single-copy sequential scenarios} \cite{Gallego14,PopescuHNL,FlaHNL,bowles2016genuinely}, where one makes a sequence of time-ordered local measurements on a single copy of the state;
    \item \textbf{multiple-copy scenarios} \cite{Palazuelos12,DaniSA,activation_zukowski,DaniNetwork,cavalcanti2012nonlocality}, where one prepares multiple copies of the state in some network structure and makes joint measurements on the local subsystems. 
\end{enumerate} 
Both scenarios are known to exhibit Bell nonlocality activation, although for generally different classes of states. One state that is commonly studied is the two-qubit isotropic state:
\begin{align}\label{isotropic}
    \rho_\alpha = \alpha\,\proj{\Phi^+}+(1-\alpha)\frac{\mathbb{1}}{4},
\end{align}

which can be seen as the result of passing one half of the two-qubit maximally entangled state $\ket{\Phi^+}=\frac{1}{\sqrt{2}}[\ket{00}+\ket{11}]$ through a depolarising channel with depolarising probability $\alpha$. The state is entangled for $\alpha>1/3$, and has an LHV model\footnote{This LHV model holds only for projective measurements, and so these examples of activation are relative to projective measurements in the standard scenario. It is generally much harder to construct LHV models for POVM measurements, although the state is known to have an LHV model for POVM measurements for $\alpha\leq 0.455$ \cite{FlaGro}. No Bell inequality violation using POVMs is known below $\alpha=1/K_3$ however, and it is generally believed that the true value is significantly higher---if not equal to $1/K_3$.} in the standard scenario if and only if $\alpha\leq 1/K_3$, where $K_3$ is Grothendieck's constant of order 3 \cite{Grothendieck,Acin2006} and the best known bounds are $0.683 < 1/K_3 < 0.697$ \cite{FlaGro,Peter}; see figure \ref{fig:main}, right. Several works have shown Bell nonlocality activation of this state. In the multiple-copy scenario, it was first shown that activation is possible for $\alpha>0.64$ \cite{activation_zukowski,DaniNetwork} by placing copies of the state in a star network topology, and later in the entire range of entanglement \cite{DaniSA} (extending \cite{Palazuelos12}) by processing many copies in a parallel scheme. In both cases one requires a large number of copies to reach a value of $\alpha$ smaller than $0.683$ (for example at least 21 states in \cite{activation_zukowski}). In the single-copy sequential scenario, no example of activation is known for the two-qubit isotropic state.

\smb

In this work, we introduce a new single-copy scenario---that we call the \emph{broadcasting} scenario---that leads to a form of Bell nonlocality activation. The scenario is inspired from the work \cite{Rafael}, where it was shown that such scenarios imply Bell inequalities that can outperform the CHSH Bell inequality for some classes of states. The causal structure of this scenario is shown in figure \ref{fig:main}, left.  The main idea is to map the bipartite state in question to a multipartite state via the application of local transformations that broadcast the local systems to a number of additional parties. The correlations arising from measurements made on the multipartite state then rule out the possibility of an LHV description for the original bipartite state. Using this scenario, we show that one can show a lack of LHV description of the state \eqref{isotropic} for values $\alpha>0.559$. For $\alpha>\frac{1}{\sqrt{3}}\approx 0.577$, we are able to prove this via the construction of a simple Bell inequality tailored to the broadcasting scenario. We also study the class of two qubit states
\begin{align}
    \rho_{\alpha,\theta} = \alpha \proj{\psi_\theta}+(1-\alpha)\rho_{A}\otimes\mathbb{1}/2,
\end{align}
where $\ket{\psi_\theta}=\frac{1}{\sqrt{2}}[\cos\theta \ket{00}+\sin\theta \ket{11}]$ and $\rho_A=\Tr_B[\proj{\psi_\theta}]$. By numerically optimising the violation of our inequality, we obtain values $\alpha, \theta$ for which $\rho_{\alpha,\theta}$ is Bell nonlocal in the broadcast scenario; see figure \ref{fig:mafalda}. 

\smb

Our results hold under the assumption that any hidden variables used to describe the full experiment obey the no-signalling principle, which is a common assumption in definitions of multipartite nonlocality \cite{MNLdefs}. This is required because one needs to consider the possibility that the transformation device broadcasts additional hidden variables, which may not be LHVs, but may exhibit nonlocal correlations (see also figure \ref{fig:DAG}).
%In order to prove a lack of hidden variable description for the original bipartite state, we require that any underlying hidden variable model for the multipartite state after the broadcasting operation state obey the no-signalling principle.  \joe{could elaborate more here.} 
Phrased more precisely, our examples of Bell nonlocality prove that in any no-signalling hidden variable theory reproducing quantum theory, those variables that describe the bipartite state in question cannot be LHVs. This is an interesting subtlety that does not arise in the other scenarios of activation that we elaborate on in section \ref{scenario}. 

\smb

Bell nonlocality is also closely linked to protocols of entanglement certification. In particular, the observation of a Bell inequality violation immediately implies that the underlying quantum state is entangled. Furthermore, since all that is required to observe a Bell inequality violation is the output statistics of the experiment, this implies that entanglement can be certified even while treating the local measurement devices as black boxes; a desirable property from both a practical and cryptographic perspective. Such protocols are consequently called \emph{device-independent} (DI), and a proof of Bell nonlocality is a DI certification of the entanglement of the state. This property carries over to the broadcasting scenario as well, and our results can equivalently be viewed as new protocols for DI entanglement certification. Moreover, by assuming that the devices are limited to preparing quantum resources---a standard assumption in DI entanglement certification---we are able to device-independently certify the entanglement of the state \eqref{isotropic} for all $\alpha>\frac{1}{2}$. This is significantly lower than the best known bounds of $1/\sqrt{2}$ (using CHSH) or $0.697$ \cite{Peter} (using 90 local measurements for each party) in the single copy regime, and we thus expect our results to be generally useful for the design of DI protocols that can tolerate high levels of noise in the state. 

\smb

\section{Preliminaries}\label{prelims}

We give here a brief recap of Bell's theorem, Bell-nonlocal and Bell-local entangled states, Bell nonlocality activation scenarios and DI certification. Readers that are familiar with these concepts may want to skip to section \ref{scenario}.  

\subsection{Bell nonlocality}

Consider two distant observers, called Alice and Bob, sharing a physical system described by variable $\lambda$ (which can be any mathematical object). Alice can interact with her part of the system by performing different measurements, which we label by an integer: $x = 0,1,... $. We also label the possible outcomes with an integer $a= 0,1, ...$ \footnote{We consider a scenario with finitely many measurements and results for convenience, note however that the analysis applies to continuous scenarios too.}. Similarly, Bob can perform measurements labelled by $y = 0,1,...$ and get outcomes $b = 0,1,...$. Upon repeating the experiment many times, Alice and Bob gain access to the joint statistics $p(ab|xy)$, that is, the probability of obtaining \emph{outcomes} $a$ and $b$, given \emph{inputs} $x$ and $y$. We call the joint statistics $p(ab|xy)$ a \emph{behaviour}.

\smb

We say that a behaviour $p(ab|xy)$ is \emph{Bell-local} if it admits the following decomposition

\begin{align}  \label{Belllocal}
p(ab|xy) = \int \Pi(\lambda) \; p_A(a|x,\lambda ) \; p_B(b|y,\lambda) \; d\lambda,
\end{align}
where $\Pi(\lambda)$ is a probability density function, and $p_A (a|x,\lambda )$, $p_B (b|y,\lambda) $ are probability distributions (for all $x,y$ and $\lambda$).
That is, the statistics can be explained by assuming the shared variable $\lambda$---called a local hidden variable (LHV)---distributed with density $\Pi(\lambda)$, and local probabilities of obtaining outcomes $a$, respectively $b$, that are independent of distant choice of input $y$, respectively $x$. A decomposition \eqref{Belllocal} is called a \emph{local hidden variable model} (LHV model) of the behaviour $p(a,b\vert x,y)$, and captures those behaviours that can be reproduced using classical shared randomness alone. If a behaviour $p(ab|xy)$ cannot be written in form \eqref{Belllocal}, we say that it is \emph{Bell-nonlocal}.

\subsubsection{Bell's theorem}

Bell's theorem states that quantum theory cannot be described by an LHV model; i.e. there exist behaviours in quantum theory that are Bell-nonlocal. More precisely, let Alice and Bob share an entangled quantum state $\rho$. Alice performs measurements from a set $\{A_{a|x} \}$ ($A_{a|x}\geq 0$ and $\sum_a A_{a|x} = \mathbb{1}$), and Bob performs measurements from a set $\{B_{b|y}\}$ (with similar conditions). The resulting behaviour is given through the Born rule by

\begin{align} \label{pQ}
p(ab|xy) = \Tr ( A_{a|x} \otimes B_{b|y} \; \rho ).
\end{align}
Bell's theorem states that there exist entangled states $\rho$ and sets of local measurements $A_{a|x}$, $B_{b|y}$ such that the behaviour $p(ab|xy)$ is Bell-nonlocal. This is typically shown with the use of a \emph{Bell inequality}; a function on behaviours that is bounded for all LHV theories, but can be violated by a quantum behaviour. We give an example below. 

\subsubsection{The CHSH inequality} \label{sec:CHSH}

The simplest Bell inequality is the CHSH inequality
\begin{align}\label{chsh}
    \langle A_0 B_0 \rangle + \langle A_0 B_1 \rangle + \langle A_1 B_0\rangle - \langle A_1 B_1 \rangle \leq 2
\end{align}{}
where correlators $\langle A_x B_y \rangle  $ are defined as 

\begin{align}
    \langle  A_x B_y \rangle   = p(00|xy) - p(01|xy) - p(10|xy) + p(11|xy),
\end{align}
that is, the expected value of the product $A \cdot B$, where $A=+1$ for $a=0$ , $A=-1$ for $a=1$, and similarly for $B$. 

\smb

In order to prove the bound of 2, note that since \eqref{chsh} is linear in the probabilities, it is enough to consider only the extremal behaviours of the set \eqref{Belllocal}. It is well known (see e.g.\ \cite{BrunnerReview} section II.B.1) that these behaviours take the form of deterministic distributions 
\begin{align}\label{det}
    p(ab\vert xy)=D_A(a\vert x)D_B(b \vert y)
\end{align}
where $D_A(a\vert x)$ is a conditional probability distribution such that $D(a \vert x)\in \{0,1\}$ $\forall a,x$, and similarly for $D_B$. 

% This can be seen as follows. Consider a arbitrary conditional probability distribution $p(a\vert x,\lambda)$. Any such distribution can be written

% \begin{align}
%     p(a\vert x,\lambda)=\int_0^1 \text{d}\mu \,D(a\vert x,\lambda,\mu)
% \end{align}
% %
% where $D(a\vert x,\lambda,\mu)=1$ if $\sum_{\tilde{a}<a}p(\tilde{a}\vert x,\lambda)\leq \mu \leq \sum_{\tilde{a}\leq a}p(\tilde{a}\vert x,\lambda)$ and zero otherwise. The distribution $D(a\vert x,\lambda,\mu)$ is deterministic, and the variable $\mu$ can be understood as an additional random variable that is used to simulate the indeterminism of the distribution $p(a\vert x,\lambda)$. Applying this process to the distributions $p_A(a\vert x,\lambda)$ and $p_B(b\vert y,\lambda)$ in \eqref{Belllocal} one obtains
% %
% \begin{align}  
% p(ab|xy) = \int_\lambda\int_{0}^1\int_0^1 \Pi(\lambda) \; D_A(a|x,\lambda,\mu_A ) \; D_B(b|y,\lambda,\mu_B) \; d\lambda\,d\mu_A\, d\mu_B.
% \end{align}
% %
% If we define a new LHV $\gamma=(\lambda,\mu_A,\mu_B)$ distributed with density 
% $\Pi(\gamma)=\Pi(\lambda)$, we can write
% \begin{align}  
% p(ab|xy) = \int_{\gamma}\Pi(\gamma) \; D_A(a|x,\gamma ) \; D_B(b|y,\gamma) \; d\gamma.
% \end{align}
% %
% Since $\Pi(\gamma)$ is a probability density, the extremal points of the set are thus \eqref{det}, of which there are a finite number, and the set is a polytope. 

For these behaviours it follows that $\langle A_x B_y \rangle = \langle A_x \rangle \langle B_y \rangle$, where 
\begin{align}
    \langle A_x \rangle=D_A(0\vert x) - D_A(1\vert x), \quad
    \langle B_y \rangle=D_B(0\vert y) - D_B(1\vert y)
\end{align}
and thus $\langle A_x \rangle, \langle B_y \rangle \in \{ -1,1\}$. One may therefore rewrite the left-hand side of \eqref{chsh} as
\begin{align}
    \langle A_0 \rangle \Big(\langle B_0 \rangle + \langle B_1 \rangle\Big) + \langle A_1\rangle \Big(\langle B_0 \rangle - \langle B_1\rangle \Big) .
\end{align}{}
Since we have $\langle B_y \rangle \in \{ -1,1\}$, it follows that one of $\langle B_0 \rangle \pm \langle B_1\rangle$ is zero, and the bound of 2 follows.

\mb

Now, taking the state $\rho = \ket{\Phi^+}\bra{\Phi^+}\ $, where $\ket{\Phi^+} = 1/\sqrt{2} ( \ket{00} + \ket{11})$, and measurements in direction $\sigma_z$, $\sigma_x$ for Alice, $\sigma_z+\sigma_x$, $\sigma_z-\sigma_x$ for Bob, a textbook calculation finds

\begin{align}
   \langle  A_0 B_0 \rangle   + \langle  A_0 B_1 \rangle   + \langle  A_1 B_0 \rangle   - \langle  A_1 B_1 \rangle    = 4 \cdot 1/\sqrt{2} = 2 \sqrt{2} .
\end{align}
Ensuring that the underlying behaviour cannot be reproduced by any LHV model.

\subsubsection{Bell-local quantum states}
A quantum state that is capable of producing Bell-nonlocal behaviours is said to be a \emph{Bell-nonlocal state}. It is easy to see that such states must be entangled. Consider a generic separable state:
\begin{align}
    \rho = \int \Pi(\lambda) \sigma^A_\lambda\otimes\sigma^B_\lambda \text{d}\lambda.
\end{align}
Performing local measurements on this state leads to 
\begin{align}\label{seploc}
    p(ab | xy) = \int \Pi(\lambda)\Tr(A_{a\vert x}\sigma^A_\lambda)\Tr(B_{b\vert y}\sigma^B_\lambda) \text{d}\lambda
\end{align}
which is of the form \eqref{Belllocal} and therefore admits an LHV model. A natural question is thus: is every entangled state Bell nonlocal? 

This question was first posed by Werner and answered in the negative \cite{Werner}. Werner presented a class of entangled mixed states---now known as Werner states---and proved that all behaviours arising from local projective measurements on the states admit an LHV model. This was later extended to a model for all POVM measurement by Barrett \cite{Barrett02}, and later extended to other classes of entangled states \cite{mafalda,bowles_sufficient,hirsch2016algorithmic,cavalcanti2016general,lhvreview,jevtic2015einstein}. Such states are known as \emph{Bell-local} states. 

The archetypal example of a Bell-local state is the two-qubit isotropic state
\begin{align}\label{isotropic2}
    \rho_\alpha = \alpha\,\proj{\Phi^+}+(1-\alpha)\frac{\mathbb{1}}{4}.
\end{align}
which is local unitary equivalent to a two-qubit Werner state. This state is entangled for $\alpha>\frac{1}{3}$ and known to be Bell-local for projective measurements for $\alpha \leq 0.683 $. For POVM measurements, the state is known to be Bell-local for $\alpha \leq 0.455$ \cite{FlaGro}, although  no example of Bell nonlocality for $\alpha<0.683$ using POVM measurements is known and this bound is thus not expected to be tight. The state is known to be Bell nonlocal for $\alpha > 0.697$ \cite{Peter} (see also figure \ref{fig:main}). 

\subsubsection{Activation of Bell nonlocality}
Some years after the result of Werner it was noticed that by subjecting some Bell-local quantum states to more complex measurement procedures, one could nevertheless prove a lack of LHV description of the state. This phenomenon is generally known as Bell nonlocality \emph{activation}. The known examples of Bell nonlocality activation can be grouped into two scenarios. 

\smb

The first group we call the \emph{single-copy sequential} scenario. Here, one performs sequences of measurements on the local subsystems, which obey a time-ordered causal structure (see \cite{Gallego14} for a detailed discussion of these scenarios). The first example of Bell nonlocality activation, shown by Popescu \cite{Popescu92}, belongs to this scenario and is known as hidden nonlocality. Here, the parties apply local filters to $\rho$, that is, a local stochastic quantum channel given by the Kraus operators $K_A, K_B$, such that the state 
\begin{align}\label{hnlstate}
    \frac{K_A\otimes K_B \; \rho \; K_A^\dagger \otimes K_B^\dagger} {\Tr[ K_A\otimes K_B \; \rho \; K_A^\dagger \otimes K_B^\dagger]}
\end{align}
violates a Bell inequality. This stochastic map can be realised by performing local measurements with measurement operators $\{K_A^\dagger K_A, \mathbb{1}-K_A^\dagger K_A\}$ and $\{K_B^\dagger K_B, \mathbb{1}-K_B^\dagger K_B\}$, where one obtains the state \eqref{hnlstate} upon obtaining the first outcomes of each measurement. One can then perform the measurements that violate the Bell inequality in a second measurement round, where the violation will be observed given the correct outputs in the first measurement round. In this way, hidden nonlocality can be seen as a subset of the possible strategies in the sequential scenario, where the Bell inequality violation from the latter measurements in the sequence precludes the existence of an LHV model for the full sequence. Popescu's method was able to activate a subset of Werner states of local dimension greater than 5 that are Bell-local for all projective measurements. This result was later strengthened by Hirsch et.\ al.\ \cite{FlaHNL} to apply to POVM-local states, and then by Bowles et.\ al.\ \cite{bowles2016genuinely} to the general multipartite scenario. Other notable examples of hidden nonlocality can be found in \cite{Gisin96,Ducuara_2016,Tendick_2020,Wang20,Pramanik_2019}. To date, all known examples of activation in the single-copy regime can be understood as hidden nonlocality, and it is an open question whether stronger results are possible by exploiting the full sequential structure. It is known however that there exist entangled states that do not have hidden nonlocality \cite{FlaNoHNL}. In particular, the isotropic state \eqref{isotropic2} was shown to display no hidden nonlocality for $\alpha \leq 0.36$. In fact, no example of Bell nonlocality activation is known for this state in the single-copy regime, even considering arbitrary sequences of measurements.

\smb

The second scenario for Bell nonlocality activation we call the \emph{multiple-copy} scenario. Here, one proves a lack of LHV description for a state by performing measurements on more than one copy of the state and exploiting the possibility to make joint measurements between the local subsystems of each copy. The usefulness of using multiple copies of a state in this context was first pointed out by Peres \cite{Peres96}, who showed that by using entanglement distillation one can generate Bell nonlocal statistics from many copies of a local (and distillable) state. With respect to nonlocality, it was shown that processing many copies of a state can increase CHSH violation \cite{Masanes06}. It was later proved that there exist states $\rho$ not violating CHSH, such that $\rho \otimes \rho$ does violate it \cite{Miguel11}. 

\smb

The first examples of multiple-copy activation were given by Aditi  et.\ al.\ \cite{activation_zukowski} and Cavalcanti et.\ al.\ \cite{DaniNetwork}, where one arranges the multiple copies of the state in a star-shaped network. Using this method, one can show activation of the state \eqref{isotropic2} for $\alpha>0.64$. A later breakthrough in the activation of nonlocality in the multiple-copy scenario was made by Palazuelos \cite{Palazuelos12}, who showed examples of Bell nonlocality activation by processing many copies of a state in parallel (as in figure \ref{fig:scenarios}). Building on Palazuelos' result, a criterion for activation of nonlocality in the same scenario was exhibited in \cite{DaniSA}. Namely, all states useful for teleportation (or equivalently states with an entanglement fraction $>1/d$, where $d$ is the local Hilbert space dimension) can be activated in the multiple-copy regime. Applied to the state \eqref{isotropic2}, these examples prove that the state can be activated in the entire range of entanglement.

We note that by combining the sequential and multiple-copy scenarios, one can show Bell nonlocality activation of any state that is distillable. This follows from two facts: (i) The existence of an entanglement distillation protocol implies a local stochastic map that maps multiple copies of the state arbitrarily close to the maximally entangled state of two qubits; (ii) The maximally entangled state violates the CHSH Bell inequality and so the multiple-copy state exhibits hidden nonlocality. In particular, since all entangled two-qubit states are distillable, this implies all such states exhibit Bell nonlocality in the multiple-copy sequential scenario. 

\subsection{Device-independent entanglement certification}
Bell nonlocality is closely linked to protocols of DI entanglement certification \cite{bancal2014device,DIENT_flavio,DIENTbarreiro,Moroder2013,Bowles2018}. From \eqref{seploc}, it follows that separable states always lead to Bell-local correlations. The observation of a behaviour $p(ab | xy)$ that is Bell-nonlocal therefore implies that the underlying state is entangled. Bell nonlocality can thus be used as a means for entanglement certification which furthermore is device-independent since the measurement devices can be treated as black boxes. 
Bell nonlocality activation also directly implies a protocol for DI entanglement certification of the underlying state, since separable states cannot exhibit activation. As a result the best known bounds for DI entanglement certification of bipartite entangled states coincide with the best bounds for Bell nonlocality activation. In the single-copy regime, this means that DI entanglement certification of the isotropic state \eqref{isotropic2} is possible for $\alpha>0.697$ \cite{Peter}. In section \ref{sec:dientcert} we show how one can in fact go beyond these bounds exploiting the quantum structure of the experiment.

\begin{figure}
    \centering
    \includegraphics[scale=0.9]{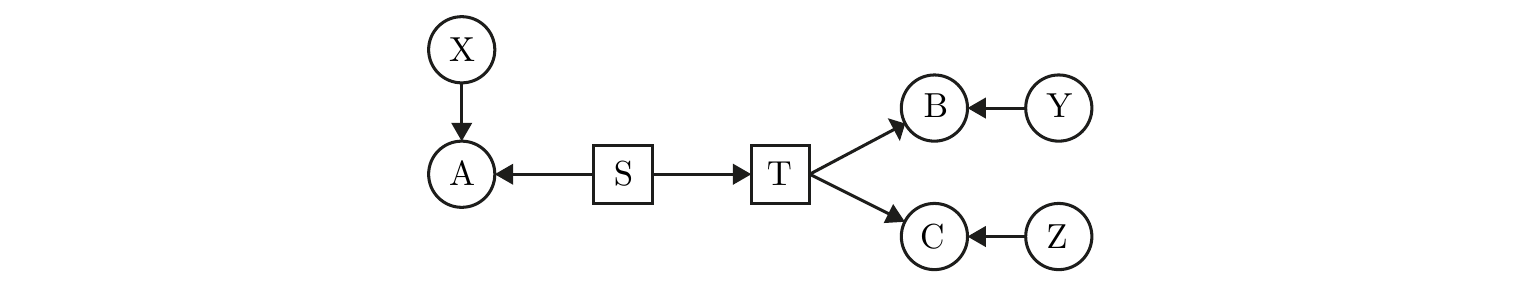}
    \caption{The causal model corresponding to the simplest broadcasting scenario. A source $S$ produces a bipartite system, one half of which is sent through a transformation $T$ that broadcasts it to two separated parties. Measurements labelled $x,y,z$ with outcomes $a,b,c$ are preformed on the resulting tripartite system. The resulting behaviour $p(a,b,c\vert x,y,z)$ is then used to rule out a local hidden variable description of the source. \label{fig:DAG}}
\end{figure}

\section{The broadcasting scenario} \label{scenario}
We now present the broadcasting scenario, focusing on the specific scenario of figure \ref{fig:DAG} for clarity. A source, denoted $S$, produces a bipartite system. One of these systems undergoes a transformation, denoted $T$, that broadcasts this system to a number of additional parties (in this case, two). Classical inputs $x,y,z$ are given to the parties (called Alice, Bob and Charlie), that return outcomes $a,b,c$, resulting in the behaviour $p(abc\vert xyz)$. In the quantum mechanical setting, the source corresponds to some bipartite quantum state $\rho_{AB_0}$, and the transformation to a completely positive trace preserving (CPTP) quantum channel $\Omega_{B_0\rightarrow BC}$ that enlarges the $B_0$ space to a tensor product space of $B$, $C$. The resulting tripartite state is thus
\begin{align}
    \rho_{ABC}=\mathbb{1}\otimes\Omega_{B_0\rightarrow BC}[\rho_{AB_0}].
\end{align}
The final behaviour obtained by making local measurements on this state is
\begin{align} \label{pRaf}
p(abc|xyz) = \Tr ( A_{a|x} \otimes B_{b|y} \otimes C_{c|z} \;  \rho_{ABC} ).
\end{align}
The question we pose is the following. \emph{Can the behaviour \eqref{pRaf} be explained by replacing the state $\rho_{AB_0}$ with local hidden variables?} A negative reply to this question would imply a type of Bell nonlocality of the state $\rho_{AB_0}$. 

To this end we consider again the causal structure of figure \ref{fig:DAG}, where the source produced LHVs  $\lambda$ distributed with density $\Pi(\lambda)$. Alice's outcome is governed by a local response function $p(a\vert x\lambda)$. The transformation device receives $\lambda$ and sends some system $\tau_\lambda$ to Bob and Charlie that they use to produce outcomes according to the joint probability distribution $p_{BC} (bc|yz,\tau_{\lambda})$. The behaviour therefore admits a decomposition
\begin{align}\label{model_gen}
p(abc\vert xyz)= \int \Pi(\lambda) \; p_A (a|x,\lambda ) \; p_{BC} (bc|yz,\tau_{\lambda})  \; d\lambda.
\end{align}
At this point, one arrives at a subtlety. Namely, what constraints does one put on the distribution $p_{BC} (bc|yz,\tau_{\lambda})$? Equivalently, what systems $\tau_\lambda$ can the transformation device prepare?

\smb

One natural option is to demand that, like the source, the transformation device produce another LHV $\lambda'$ that depends on $\lambda$ to send to Bob and Charlie. That is, $\tau_\lambda$ is an LHV $\lambda'(\lambda)$ and $p_{BC} (bc|yz,\tau_\lambda)$ therefore factorises as $ p_B (b|y,\lambda'(\lambda)) p_C (c|z,\lambda(\lambda'))$, leading to\footnote{Note that we have assumed that $\lambda'$ is a deterministic function of $\lambda$. In principle, the transformation device could act stochastically and produce a new LHV $\lambda'$ conditioned on $\lambda$ described by some conditional probability distribution $q(\lambda'\vert\lambda)$. By defining a new LHV $\mu=(\lambda,\lambda')$ distributed with density $\Pi(\lambda)q(\lambda'\vert \lambda)$, this in-determinism can be absorbed by the source, so that the distribution admits the decomposition \eqref{model1b} with $\lambda$ replaced by $\mu$. Thus, the form \eqref{model1b} is general.}
\begin{align}  \label{model1}
p(abc|xyz) &= \int \Pi(\lambda) \; p_A (a|x,\lambda ) \; p_B (b|y,\lambda'(\lambda)) p_C (c|z,\lambda'(\lambda))  \; d\lambda\\
&= \int \Pi(\lambda) \; p_A (a|x,\lambda ) \; \tilde{p_B} (b|y,\lambda) \tilde{p_C} (c|z,\lambda))  \; d\lambda . \label{model1b}
\end{align}
Note that this is simply the standard definition of Bell nonlocality for a three-party scenario. Suppose now that one sees a violation of $\eqref{model1b}$. This implies Bell nonlocality somewhere due to a lack of LHV model for the three parties. The problem, however, is that one cannot conclude that this nonlocality originated from the original state $\rho_{AB_0}$. Consider for instance the following experiment. The source prepares a state $\rho_{AB_0}$, however the transformation device ignores the $B_0$ subsystem and instead creates a fresh maximally entangled state which it sends to Bob and Charlie (i.e.\ if $B_0$ is a qubit space, the Kraus operators for the channel are $\ketbra{\Phi^+}{0}$ and $\ketbra{\Phi^+}{1}$). Bob and Charlie now hold a maximally entangled state and can violate the CHSH inequality. The resulting behaviour will not admit a decomposition \eqref{model1b} regardless of what Alice measures. This will be the case for any choice of $\rho_{AB_0}$, including separable $\rho_{AB_0}$. Thus, although a violation of \eqref{model1b} implies some Bell nonlocality, one cannot rule out a  LHV description of $\rho_{AB_0}$, since the nonlocality may have originated from the transformation device alone. \smb

To make a statement about $\rho_{AB_0}$, we thus need to consider models that rule out the above possibility. To do this, we allow the systems $\tau_\lambda$ to be arbitrary non-signalling resources, so that the distribution $p_{BC} (bc|yz,\tau_\lambda)$ is any normalised joint conditional probability distribution that obeys the no-signalling constraints: 
\begin{align} \label{NScons}
    \sum_b p_{BC} (bc|yz,\tau_\lambda) &= \sum_b p_{BC} (bc|y'z,\tau_\lambda) \;\forall y,y',c,z,\tau_\lambda, \\ \label{NScons2}
     \sum_c p_{BC} (bc|yz,\tau_\lambda) &= \sum_c p_{BC} (bc|yz',\tau_\lambda) \;\forall z,z',b,y,\tau_\lambda. 
\end{align}
Thus, the full distribution can be decomposed
\begin{align}  \label{modelNS}
p(abc|xyz) = \int \Pi(\lambda) \; p_A (a|x,\lambda ) \; p^{NS}_{BC} (bc|yz,\lambda)  \; d\lambda .
\end{align}
where the $p^{NS}_{BC} (b,c|y,z,\lambda)$ are arbitrary non-signalling distributions satisfying \eqref{NScons}, \eqref{NScons2} between Bob and Charlie, for each value $\lambda$. Note that for a finite number of inputs and outcomes, the sets of possible distributions $p_A (a|x,\lambda )$ and $p^{NS}_{BC} (bc|yz,\lambda)$ are both convex polytopes, which follows from the fact that they are defined by a finite number of linear inequalities. There are therefore a finite number of extremal distributions in each of these sets. In particular the extremal $p_A (a|x,\lambda )$ correspond to deterministic distributions, and the extremal $p^{NS}_{BC} (bc|yz,\lambda)$ to the (possibly non-deterministic and Bell non-local) extremal points of the no-signalling set (see e.g. \cite{BrunnerReview}). The set \eqref{modelNS} can thus be equivalently defined 
\begin{align}  \label{modelNSextreme}
p(abc|xyz) = \sum_\lambda p(\lambda_1,\lambda_2) \; p_A (a|x,\lambda_1 ) \; p^{NS}_{BC} (bc|yz,\lambda_2),
\end{align}
where $p_A (a|x,\lambda_1 )$ and $p^{NS}_{BC} (bc|yz,\lambda_2)$ are these extremal distributions, and is therefore a convex polytope.

Assuming the no-signalling condition, any violation of \eqref{modelNS} cannot originate from the transformation device alone, so long as it produces non-signalling resources (as is the case for quantum theory and the majority of GPT theories). One can thus make the following statement.\smb  \emph{In any no-signalling hidden variable theory reproducing quantum theory, a violation of \eqref{modelNS} implies that those variables describing the state $\rho_{AB_0}$ cannot be LHVs.}\smb

We thus take the set of behaviours admitting a decomposition \eqref{modelNS} as our analogue of the local set in the broadcasting scenario. If a state leads to behaviours that violate \eqref{modelNS}, we say it is \emph{broadcast-nonlocal}. 

\smb

\section{Activation of two-qubit states}\label{sec:activation}
We now study Bell nonlocality activation of the isotropic state \eqref{isotropic2} in the broadcasting scenario. As a main result, we show analytically that the state is broadcast-nonlocal for $\alpha>1/\sqrt{3}$. To find this result we employed the convex optimisation methods in section \ref{sec:numerics} and the heuristic algorithm described in appendix \ref{ap:heuristic}. A slightly lower value of $\alpha=0.559$ was found with the same methods, although we failed to convert this strategy to a nice analytic form (see auxiliary files for the numerical files needed to verify this result). 

\subsection{Simple inequality for three parties}
To prove our results, we construct a correlation inequality that is satisfied by all behaviours admitting a decomposition \eqref{modelNS}. Consider a scenario in which Alice has 3 inputs $x=0,1,2$, and Bob and Charlie both have 2 inputs $y=0,1$, $z=0,1$. All outcomes are dichotomic with $a,b,c=\pm 1$. Define the three body correlators
\begin{align}\label{3corr}
\langle  A_x B_y C_z \rangle = \sum_{a,b,c=\pm 1} a b c \; p(abc\vert xyz),
\end{align}
the two body correlators for Alice and Bob
\begin{align}\label{2corr}
\langle A_x B_y \rangle = \sum_{a,b=\pm 1} ab\; p(ab\vert xy)
\end{align}
(and similarly for other pairs of parties), and the single body correlators for Alice
\begin{align}\label{1corr}
\langle A_x \rangle = \sum_{a=\pm 1} a\;p(a\vert x)  
\end{align}
(and similarly for Bob and Charlie). The inequality reads
\begin{multline}\label{ineq_1Alice}
    \mathcal{I}=\langle  A_0 B_0 C_0 \rangle   + \langle  A_0 B_1 C_1 \rangle   + \langle  A_1 B_1 C_1 \rangle   - \langle  A_1 B_0 C_0 \rangle \\  + 
    \langle  A_0 B_0 C_1 \rangle   + \langle  A_0 B_1 C_0 \rangle   + \langle  A_1 B_0 C_1 \rangle   - \langle  A_1 B_1 C_0  \rangle \\ -2 \langle  A_2 B_0 \rangle   + 2 \langle  A_2 B_1 \rangle   \leq 4,
\end{multline}
which can be shown to be a facet of the broadcast-local polytope. Note that with a slight abuse of notation the above can be written
\begin{align}\label{ineq_chsh}
    \mathcal{I}=\langle \text{CHSH}[A_0,A_1,C_0,C_1](B_0+B_1)\rangle +\langle A_2(B_1-B_0)\rangle \leq 4,
\end{align}
where $\text{CHSH}[A_0,A_1,C_0,C_1]= A_0C_0+A_0C_1-A_1C_0+A_1C_1$ is a relabelling of the well known CHSH Bell inequality \cite{CHSH}. We first prove the bound of $4$ for behaviours \eqref{modelNS}. 

\mb

\begin{proof}
Since \eqref{ineq_1Alice} is linear in the probabilities, we can restrict ourselves to extremal points of the set \eqref{modelNS}. These points take the form 
\begin{align}\label{factorpoints}
    p(abc\vert xyz)=D(a\vert x)P(bc\vert yz)
\end{align}
where $D$ is a deterministic behaviour such that $D(a\vert x)\in\{0,1\}\;\forall a,x$, and $P$ is an extremal behaviour of the 2 parties, 2 inputs, 2 outputs scenario. For the points \eqref{factorpoints}, it follows that the three body correlators factorise
\begin{align}
    \langle  A_x B_y C_z \rangle = \langle A_x \rangle \langle B_y C_z \rangle.
\end{align}
We thus have 
\begin{multline}
    \mathcal{I}=\Big( \langle A_0 \rangle - \langle A_1 \rangle \Big) \Big(\langle B_0C_0\rangle+\langle B_1C_0 \rangle \Big)\\ +\Big(\langle A_0 \rangle + \langle A_1\rangle\Big)\Big( \langle B_0C_1 \rangle+\langle B_1C_1 \rangle\Big) \\ + 2\langle A_2\rangle   \Big(\langle B_1\rangle -\langle B_0\rangle\Big).
\end{multline}
Since Alice's strategy is deterministic, one has $\langle A_x \rangle =\pm 1$ and thus either the first or the second term of the above must be equal to zero. Due to the symmetry of the inequality, we can take this to be the second term without loss of generality. Using $\vert\langle A_x \rangle\vert\leq 1$ we thus have
\begin{align}
    \mathcal{I}&\leq 2\vert\langle B_0C_0\rangle+\langle B_1C_0 \rangle \vert+ 2 \vert \langle B_1\rangle -\langle B_0 \rangle \vert. \label{ineqabs}
\end{align}
We now consider the extremal behaviours $P(bc\vert yz)$ for Bob and Charlie. These are known to be of two types \cite{barrett2005nonlocal}: (i) local deterministic behaviours and (ii) local relabellings of the PR box. In case (i), the correlators $\langle B_y C_z \rangle$ factorise as $\langle B_y \rangle \langle C_z \rangle$ and one has 
\begin{align}
    \mathcal{I}\leq 2\vert \big( \langle B_0\rangle + \langle B_1 \rangle \big) \langle C_0 \rangle \vert + 2 \vert\langle B_1\rangle -\langle B_0\rangle\vert\leq 4
\end{align}
since either $\langle B_0\rangle + \langle B_1 \rangle$ or $\langle B_0\rangle - \langle B_1 \rangle$ is zero. In case (ii), since the PR box is such that $\langle B_0 \rangle = \langle B_1 \rangle =0$, from \eqref{ineqabs} one also obtains the bound 4. 
\end{proof}

\subsubsection{Quantum violation}
The inequality \eqref{ineq_1Alice} can be violated by the isotropic state \eqref{isotropic2}. Let Alice measure the triple of anti-commuting qubit observables 
\begin{align}
    A_0 = \sigma_z, \;
    A_1 = \sigma_x, \;
    A_2 = \sigma_y.
\end{align} 
Charlie and Bob measure the observables  
\begin{equation}
\begin{aligned}
    C_0 &= \sigma_z, \; C_1 =\sigma_x,   \\
    B_0 &= \cos(\phi) \sigma_x + \sin(\phi) \sigma_y, \; B_1=\cos(\phi) \sigma_x - \sin(\phi) \sigma_y,
    \end{aligned}
\end{equation}
with $\tan(\phi)=1/\sqrt{2}$. The channel $\Omega_{B_0\rightarrow BC}$ is an isometry given by
\begin{align}\label{isometry}
    U=\ket{\psi_0}\bra{0}+\ket{\psi_1}\bra{1}
\end{align}
with 
\begin{align}
    \ket{\psi_0}&=\sin\beta \ket{\Phi^-} + \cos\beta\ket{\Psi^+} \\
    \ket{\psi_1}&=-\cos\beta \ket{\Phi^-} + \sin\beta\ket{\Psi^+},
\end{align}
and where $\ket{\Phi^-}$, $\ket{\Psi^+}$ are the usual Bell states and $\beta=\pi/8$. Taking the isotropic state, one finds via direct calculation that 
\begin{align}
    \mathcal{I}=\alpha \; 4\sqrt{3}.
\end{align}
Thus, for $\alpha>1/\sqrt{3}$ one obtains a violation.

\begin{figure}
\centering
\begin{tikzpicture}[scale=0.8]
\begin{axis}[ 
xlabel={$\theta$}, 
ylabel={visibility $\alpha$},
xmin=0,
xmax= 0.7854,
%ymin=0.2,
%ymax=1,
%legend pos = north west
] 
\pgfplotstableread{ResMafalda_.dat}\loadedtable;
\addplot+[mark options={scale=0.3,mark=*},line width = 1.2pt] table {\loadedtable};
%\addplot [line width=1pt] {1/3};
\pgfplotstableread{ResMafalda_chsh.dat}\loadedtable;
\addplot+[mark options={scale=0.3,mark=none},line width = 1.2pt] table {\loadedtable};
\legend{Broadcast-nonlocal,CHSH-nonlocal};
\end{axis}
\end{tikzpicture}
\caption{Critical value of the visibility $\alpha$ above which the state \eqref{mafalda} is broadcast-nonlocal (blue line), and Bell nonlocal using the CHSH inequality in the standard scenario (red line). The blue curve was obtained by numerically optimising the violation of inequality \eqref{ineq_1Alice}. \label{fig:mafalda}}
\end{figure}
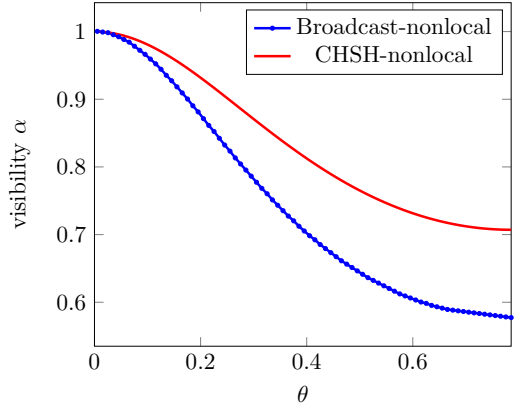

We have also investigated the violation of inequality \eqref{ineq_1Alice} for the larger class of two-qubit states
\begin{align}\label{mafalda}
    \rho_{\alpha,\theta} = \alpha \proj{\psi_\theta}+(1-\alpha)\rho_{A}\otimes\mathbb{1}/2,
\end{align}
where $\ket{\psi_\theta}=\frac{1}{\sqrt{2}}[\cos\theta \ket{00}+\sin\theta \ket{11}]$ and $\rho_A=\Tr_B[\proj{\psi_\theta}]$. This state is entangled for $\alpha>1/3$ for all $\theta>0$. 
We plot the critical visibility at which the Bell inequality \eqref{ineq_1Alice} is no longer violated, and compare this to the equivalent value for the widely used CHSH Bell inequality \cite{CHSH} in the standard scenario. As with the isotropic state, the inequality \eqref{ineq_1Alice} allows for lower visibilities than the CHSH inequality for all values of $\theta$, see figure \ref{fig:mafalda}. Unlike the isotropic state however, it is not known if there exists an LHV model for the state \eqref{mafalda} in the region above the broadcast-nonlocal bound for $\theta \neq \pi/4$, and so it is not known if these values of the visibility correspond to activation or not. Given that this is the case for $\theta=\pi/4$ however, we expect there to be some activation, at least for a range of $\theta$ close to $\pi/4$.

\mb

\subsection{Symmetric four-partite inequality} \label{sec:symmetric4}
\begin{figure}
    \centering
    \includegraphics[scale=0.9]{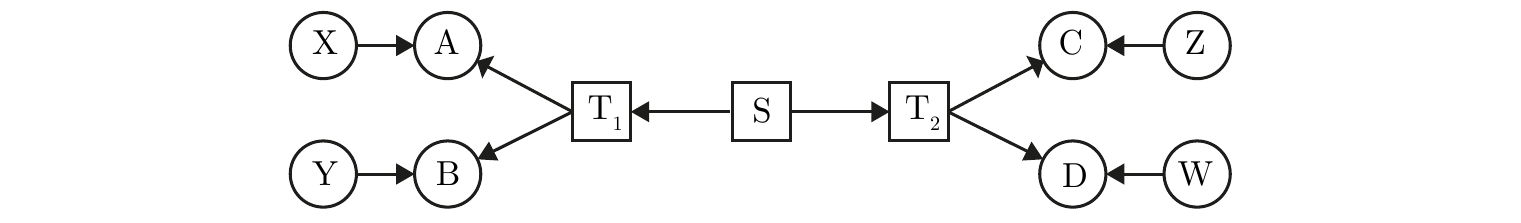}
    \caption{A four-partite broadcast scenario. Transformations are applied to both subsystems of the source.}
    \label{fourpartite}
\end{figure}

One can also obtain activation of the isotropic state in the same range as above in a symmetric scenario where one applies a channel to both local subsystems resulting in a 4-partite state $\rho_{ABCD}$ (see figure \ref{fourpartite}), where each party now has two inputs. Following the same logic as before, the relevant behaviours are now of the form 
\begin{align}  \label{model2b}
p(abcd|xyzw) = \int \Pi(\lambda) \; p^{NS}_{AB} (a,b|x,y,\lambda) \; p^{NS}_{CD} (c,d|z,w,\lambda)  \; d\lambda.
\end{align}
Where $p^{NS}_{AB}$ and $p^{NS}_{CD}$ are non-signalling distributions.  The inequality we use to show activation in this scenario then reads
\begin{align}\label{ineq_2Alice}
    \mathcal{I}=\langle  A_0 B_0 C_0 D_0 \rangle   + \langle  A_0 B_0 C_1 D_0 \rangle   + \langle  A_0 B_1 C_0 D_0 \rangle   + \langle  A_0 B_1 C_1 D_0 \rangle   + \nonumber\\
    -\langle  A_1 B_0 C_0 D_0 \rangle   + \langle  A_1 B_0 C_1 D_0 \rangle   - \langle  A_1 B_1 C_0 D_0 \rangle   + \langle  A_1 B_1 C_1 D_0 \rangle   - \nonumber\\
    2 \langle  B_0 D_1 \rangle   + 2 \langle  B_1 D_1 \rangle   \leq 4 .
\end{align}
We first derive the bound of $4$.

\begin{proof}
We again may focus on extremal behaviours of the set \eqref{ineq_2Alice}. Since the extremal points of non-signalling behaviours for 2 inputs and 2 outputs are either local deterministic distributions, or symmetries of the PR box behaviour \cite{barrett2005nonlocal}, the extremal behaviours of the set \eqref{ineq_2Alice} can be grouped into three groups: (i) All parties have local deterministic outcomes, (ii) A and B have local deterministic outcomes, and C and D share a PR box (or vice-versa) and (iii) Both A and B, and C and D share a PR box. In all cases, the full body correlators factorise as 
\begin{align}
    \langle  A_x B_y C_z D_w \rangle = \langle  A_x B_y \rangle\langle C_z D_w \rangle.
\end{align}
We thus write \eqref{ineq_2Alice} as
\begin{multline}\label{ineq_2Alicecond}
    \mathcal{I} = \Big(\langle A_0B_0 \rangle+ \langle A_0B_1 \rangle\Big) \Big(\langle C_1 D_0 \rangle + \langle C_0 D_0\rangle\Big) \\ + \Big(\langle A_1 B_0 \rangle + \langle A_1 B_1 \rangle \Big)\Big(\langle C_1 D_0 \rangle - \langle C_0 D_0\rangle\Big) \\ - 2 \Big([\langle B_1 \rangle -\langle B_0 \rangle] \langle D_1 \rangle\Big) .
\end{multline}
Consider the final term of the inequality. This term is non-zero if and only if neither pair uses a PR box and $\langle B_0 \rangle = - \langle B_1 \rangle$. In this case, since A and B use a deterministic strategy, one has 
\begin{align}
    \langle A_0B_0 \rangle+ \langle A_0B_1 \rangle = \langle A_0 \rangle (\langle B_0 \rangle+ \langle B_1 \rangle) = 0
\end{align}
and so the first term of \eqref{ineq_2Alicecond} is zero (and similarly for the second term). Thus, one obtains the bound $4$. For all other cases, the final term of \eqref{ineq_2Alicecond} is zero and we can focus on the first two terms. Here, it follows that for all strategies of C and D either $\langle C_1 D_0 \rangle + \langle C_0 D_0\rangle$ or $\langle C_1 D_0 \rangle - \langle C_0 D_0\rangle$ is zero, and again one obtains $\mathcal{I}\leq 4$.

\end{proof}

\subsubsection{Quantum violation}
Again we consider the isotropic state \eqref{isotropic2}. The two local channels applied to the state are identical and are again given by \eqref{isometry}. Alice and Bob's local observables are 
\begin{equation}
    A_0 = \sigma_x,\; A_1 = \sigma_z, \quad
    B_0 = (\sigma_x+\sigma_y+\sigma_z)/\sqrt{3} ,\; B_1 = (\sigma_x-\sigma_y+\sigma_z)/\sqrt{3}.
\end{equation}
Charlie's and Bob's local observables are 

\begin{equation}
    C_0 = \sigma_x, \; C_1 =\sigma_z, \quad D_0 = \sigma_x, \; D_1 =\sigma_y. 
\end{equation}
Using this strategy $\mathcal{I} =  12 \alpha /\sqrt{3} = 4\alpha \cdot 3/\sqrt{3} = 4\alpha\sqrt{3} $, thus leading to activation for $\alpha>1/\sqrt{3}$. 

\section{Device-independent entanglement certification in the broadcasting scenario}\label{sec:dientcert}
In DI entanglement certification, one aims at certifying the entanglement of the state in an experiment based solely on the observation of the experimental statistics. Here, we show how the broadcasting scenario can be used to design protocols for DI entanglement certification. To do this, we consider the four-partite scenario of figure \ref{fourpartite}. To achieve a DI certification of entanglement, one needs to show that the correlations are incompatible with the source producing any separable state, that is, one must show that the correlations are incompatible with the causal network of figure \ref{fourpartite}, where the source is replaced by a separable state. 

% An alternative possibility to the above is to demand that the systems $\tau_\lambda$ be quantum systems of arbitrary dimension. Following the same logic, the set of behaviours that are compatible with a source of LHVs is 
% %
% \begin{align}  \label{modelQ}
% p(abc|xyz) = \int \Pi(\lambda) \; p_A (a|x,\lambda ) \; p^{Q}_{BC} (bc|yz,\lambda)  \; d\lambda,
% \end{align}
% %
% where the behaviours $p^{Q}_{BC} (bc|yz,\lambda)$ are in the set of quantum correlations (a subset of the no-signalling set of correlations). This is less in the spirit of Bell nonlocality, since one assumes a quantum structure of the experiment. However, it is precisely the assumption one makes in DI entanglement certification. In section \ref{sec:dientcert} we elaborate of this idea and show how this leads to the construction of DI entanglement witnesses tailored to the broadcasting scenario. In the following section we focus on broadcast-nonlocality of the two-qubit states. 

Consider a state $\rho_{A_0C_0}$ produced by the source. This state is separable iff it admits a decomposition
\begin{align}
    \rho_{SEP}=\int \text{d}\lambda \Pi(\lambda)\sigma_\lambda^{A_0}\otimes\sigma_\lambda^{C_0}.
\end{align}
Following the local transformations, the most general four-partite state than one can obtain from this state is of the form
\begin{align}\label{bisep}
    \rho_{ABCD}=\int \text{d}\lambda \Pi(\lambda)\sigma_{\lambda}^{AB}\otimes \sigma_\lambda^{CD}
\end{align}
where $\sigma^{AB}_\lambda=\Omega_{A_0\rightarrow AB}[\sigma_\lambda^{B_0}]$ and $\sigma^{CD}_\lambda=\Omega_{C_0\rightarrow CD}[\sigma_\lambda^{C_0}]$. Performing local measurements on this state thus leads to behaviours of the form 
\begin{align}  \label{model3}
p(abcd|xyzw) &= \Tr\left[\rho_{ABCD} A_{a\vert x}\otimes B_{b\vert y} \otimes C_{c\vert z} \otimes D_{d\vert w} \right] \\
&= \int \Pi(\lambda)\Tr\left[\sigma^{AB}_\lambda A_{a\vert x}\otimes B_{b\vert y} \right] \Tr\left[ \sigma^{CD}_\lambda C_{c\vert z} \otimes D_{d\vert w} \right]\\ 
&=\int \Pi(\lambda) \; p^Q_{AB} (ab|xy\lambda) \; p^Q_{CD} (cd|zw\lambda)  \; d\lambda,
\end{align}
where $p^Q_{AB}$, $p^Q_{CD}$ are arbitrary quantum behaviours. Thus, the possible correlations correspond to those that can be obtained by a general four-partite state $\rho_{AB\vert CD}$ (of arbitrary dimension), that is separable with respect to the bipartition $AB$ vs $CD$. Thus, if we consider an expression $\mathcal{I}[p(abcd\vert xyzw)]$ in this scenario and define $\beta$ the maximum value obtained by biseperable states \eqref{bisep}, a violation of the inequality $\mathcal{I}[p(abcd\vert xyzw)]\leq \beta$ certifies that the state $\rho_{A_0C_0}$ is entangled in a DI manner. 

Before proceeding to a specific example, we clarify the assumptions of the certification. As is standard in DI certification we assume that all measurement and transformation devices act on separate locations in space and thus correspond to disjoint Hilbert spaces. Strictly speaking, the space $A_0$ should be understood to encapsulate all the degrees of freedom `to the left' of the source in figure \ref{fourpartite}; i.e. the space on which Alice's transformation device acts as well as the spaces $A$ and $B$ (and analogously for $C_0$). No assumption is needed on the causal ordering of the transformation and measurements (although it would clearly be useless to apply a transformation after the measurement process).

\smb

\subsection{Entanglement certification of isotropic states}
We now apply this reasoning to the isotropic state, where we show that DI entanglement certification is possible for $\alpha>1/2$. The expression we will use is the four-partite Klyshko-Belinskii expression \cite{belinskiui1993interference, gisin1998bell}: 
\begin{align}\label{MABK}
    \mathcal{I}=2^{-\frac{3}{2}}\sum_{\vec{x}\in\{0,1\}^4}\cos \left[ \frac{\pi}{4}(2 \vert \vec{x}\vert - 3) \right]\langle A_{x_1}B_{x_2}C_{x_3}D_{x_4} \rangle 
    %\frac{1}{4}[\expect{A_1B_0C_0D_0} - \expect{A_0B_1C_0D_0} + \expect{A_0B_0C_1D_0} + \expect{A_1B_1C_1D_0} \\ - \expect{A_0B_0C_0D_1} - \expect{A_1B_1C_0D_1} + \expect{A_1B_0C_1D_1} - \expect{A_0B_1C_1D_1}]
\end{align}
where $\vert \vec{x}\vert=\sum_{i=1}^4 x_i$. The bound $\mathcal{I}\leq 1$ for LHV theories is a well known Bell inequality in the standard four-party Bell scenario. The maximum quantum violation of the expression is $2^{3/2}$ and is achieved by the four-partite GHZ state
\begin{align}
    \ket{\text{GHZ}}=\frac{1}{\sqrt{2}}\left( \ket{0000} + \ket{1111}\right)
\end{align}
and measurements observables that lie in the $x-y$ plane of the Bloch sphere for each of the parties. In \cite{DIEWs} (using \cite{nagata}), it was proven that the $k$-producible bound of the inequality---the maximum possible value obtainable by states that contain entanglement between at most $k$ parties---is given by $\mathcal{I}\leq 2^{(k-1)/2}$. Note that the state \eqref{bisep} is $2$-producible, since there are at most two parties entangled. Thus, a violation of the inequality $\mathcal{I}\leq 2^{1/2}$ is a DI entanglement certification of $\rho_{A_0C_0}$ and \eqref{MABK} can be used as a DI entanglement witness in the broadcasting scenario.

Now, take the local channels in figure \ref{fourpartite} to be isometries given by
\begin{align}
    U=\ketbra{00}{0}+\ketbra{11}{1}.
\end{align}
With this choice, and taking $\rho_{A_0C_0}$ to be the isotropic state we obtain
\small
\begin{align}
    \rho_{ABCD}=\alpha\proj{\text{GHZ}} + \frac{1}{4}(1-\alpha)\left[\proj{0000}+\proj{0011}+\proj{1100}+\proj{1111}\right].
\end{align}
\normalsize
We now perform the measurements that give the maximum violation of $\mathcal{I}$ for the GHZ state. Note that the second term in the state will contribute zero to $\mathcal{I}$  since it is diagonal in the $z$ basis, and the measurements are chosen in the $x-y$ plane. One therefore obtains the value $\alpha\cdot 2^{3/2}$, and has a DI certification of entanglement for 
\begin{align}
    \alpha\cdot 2^{3/2} > 2^{1/2},
\end{align}
that is, for $\alpha>1/2$. We suspect that further investigation into the biseparable bounds of known Bell inequalities could lead to even stronger examples.

It may also be possible that stronger examples are already possible in the three partite scenario of figure \eqref{fig:DAG}, and in the next section we give a numerical method that can be useful to find such examples. We note that  the Bell inequality \eqref{ineq_1Alice} has the same bound of $4$ when considering quantum resources, since this bound can be achieved by a local deterministic (and thus quantum) strategy. This means that the same critical visibility of $1/\sqrt{3}$ is obtained for DI entanglement certification. 
\mb

\section{Numerical methods} \label{sec:numerics}
Here we outline two numerical methods that allow one to tackle the problem of testing compatibility with the causal networks considered in the previous sections. In the case of section \ref{sec:activation}, this can be achieved via a linear program, and the methods we present (combined with the algorithm in appendix \ref{ap:heuristic}) were used to find the results of section \ref{sec:activation}. We present two approaches to this end; the second uses a possibly unknown trick to reduce memory usage and may be of independent interest.  For DI entanglement certification, we present a method based on semi-definite programming to define a relaxation to the problem that in practice can give good results. This method works for scenarios in which the broadcasting channel is applied to only one party. It would be interesting to use this method to look for stronger examples that that presented in section \ref{sec:dientcert}.

\subsection{Linear programming methods}
Let us consider again the set of behaviours defined in \eqref{modelNS}:
\begin{align} \label{model2_2}
p(abc|xyz) = \int \Pi(\lambda) \; p_A (a|x,\lambda ) \; p^{NS}_{BC} (bc|yz,\lambda)  \; d\lambda,
\end{align}
and denote by $\mathcal{L}_{NS}$ the set of behaviours admitting the above decomposition. It is well known (see e.g. \cite{BrunnerReview}) that the sets of distributions $p_A (a|x,\lambda )$ and $p^{NS}_{BC} (bc|yz,\lambda)$ are both convex and have a finite number of extremal elements, that is, they are convex polytopes. Denote these extremal elements by $\{D_A(a\vert x, i)\}_{i}$ and $\{E_{BC}(bc|yz,j)\}_{j}$ where $i$ and $j$ index the elements. In this scenario, these elements correspond to deterministic behaviours for Alice, and the extremal behaviours of the no-signalling polytope for Bob and Charlie. Thus, there always exist non-negative weights $q(i\vert\lambda)$, $w(j\vert\lambda)$ such that $\sum_i q(i\vert\lambda)=\sum_j w(j\vert\lambda) =1$ such that 
\begin{align}\label{replace_det}
    p_A (a|x,\lambda )= \sum_{i}q(i\vert\lambda) D_A(a\vert x ,i),\quad\quad
    p^{NS}_{BC} (bc|yz,\lambda) =  \sum_{j} w(j\vert\lambda) E_{BC}(bc|yz,j).
\end{align}
Substituting this in \eqref{model2_2} we find
\begin{align} \label{model2_det}
p(abc|xyz) = \sum_{i,j} p_{ij} \; D_A(a\vert x ,i) \; E_{BC}(bc|yz,j),
\end{align}
where $ p_{ij}=\int \Pi(\lambda)q(i\vert\lambda)w(j\vert\lambda) d\lambda$. Hence, since the distributions $D_A$ and $E_{BC}$ are fixed,  the problem of deciding membership in $\mathcal{L}_{NS}$ is equivalent to finding positive numbers $p_{ij}$ such that the above is satisfied. Moreover, since the above expression is linear in $p_{ij}$ this can be cast as a linear programming instance, i.e., one solves the linear program
\begin{align}\label{LP}
    &\text{find } p_{ij}\geq 0 \;\;\text{such that} \\ \nonumber
    & p(abc|xyz) = \sum_{i,j} p_{ij} \; D_A(a\vert x ,i) \; E_{BC}(bc|yz,j) \;\;\forall a,b,c,x,y,z.
\end{align}
If the behaviour is found to be outside of $\mathcal{L}_{NS}$, the dual of the optimisation returns a corresponding separating hyperplane that one can interpret as a generalised Bell inequality for the broadcasting scenario.  

In practice, one is limited to scenarios in which the number of optimisation variables $p_{ij}$ can be stored in memory. Since the number of extremal distributions increases exponentially in the number of each input, this means that \eqref{LP} can become intractable for relatively simple scenarios. One method to partly overcome this, is to use the fact the distributions $p^{NS}_{BC} (bc|yz,\lambda)\in \mathcal{P}_{NS}$ have a linear characterisation, given in \eqref{NScons}, \eqref{NScons2}. That is
\begin{align}
    p^{NS}_{BC} (bc|yz,\lambda)\in \mathcal{P}_{NS} \iff \quad 
    &\sum_b p^{NS}_{BC} (bc|yz,\lambda) = \sum_b p^{NS}_{BC} (bc|y'z,\lambda) \;\forall y,y',c,z,\lambda, \\ 
     &\sum_c p^{NS}_{BC} (bc|yz,\lambda) = \sum_c p^{NS}_{BC} (bc|yz',\lambda) \;\forall z,z',b,y,\lambda. 
\end{align}
If in \eqref{model2_2} we replace $p_A (a|x,\lambda )$ only (through \eqref{replace_det}) then we find
\begin{align} \label{malin}
p(abc|xyz) &= \sum_i D_A (a|x,i )\left[\int \Pi(\lambda) q(i\vert \lambda) \; p^{NS}_{BC} (bc|yz,\lambda)  \; d\lambda \right] \nonumber \\
&= \sum_i D_A (a|x,i ) \,\tilde{p}^{NS}_{BC}(bc|yz,i),
\end{align}
where $\tilde{p}^{NS}_{BC}(bc|yz,i)$ is a sub-normalised non-signalling distribution with normalisation $\int \Pi(\lambda) q(i\vert \lambda)d\lambda$ that is the same for all $y,z$, and $\tilde{p}^{NS}_{BC}(bc|yz,i)\in\mathcal{P}_{NS}$ since it still satisfies the constraints \eqref{NScons}, \eqref{NScons2}. One can therefore also test membership in \eqref{model2_2} with an equivalent LP by treating the distributions $\tilde{p}^{NS}_{BC}(bc|yz,i)$ as optimisation variables:
\begin{align}\label{LP2}
    &\text{find } \tilde{p}^{NS}_{BC}(bc|yz,i) \geq 0 \;\;\text{such that} \nonumber\\[8pt] \nonumber
    & p(abc|xyz) = \sum_i D_A (a|x,i ) \tilde{p}^{NS}_{BC}(bc|yz,i), \;\;\forall a,b,c,x,y,z; \\
    &\tilde{p}^{NS}_{BC}(bc|yz,i)\in \mathcal{P}_{NS} \;\;\forall i, \quad \sum_{bc}\tilde{p}^{NS}_{BC}(bc|yz,i) = \sum_{bc}\tilde{p}^{NS}_{BC}(bc|y'z',i) \;\;\forall y,z,y',z',i.
\end{align}
The number of optimisation variables now scales with the number of extremal elements of Alice's distribution, at the expense of introducing a similar number of constraints to the optimisation. In practice, this requires less memory than \eqref{LP} and allows one to tackle significantly larger scenarios. Both linear programs can be applied to the causal structure of figure \ref{fourpartite} (or scenarios with any number of parties) by decomposing the relevant distributions as convex mixtures of their extremal elements in a similar fashion. One can also define closely related linear programs to calculate the robustness to noise to the set $\mathcal{L}_{NS}$ given a state with a linear noise parameter, and fixed measurements and channel(s); see Appendix \ref{linpr}. 

\subsection{Semi-definite programming hierarchy for entanglement certification}
Here we wish to tackle membership in scenarios where the transformation device prepares quantum systems. The approach we present is applicable to the scenario of figure \ref{fig:DAG}. Denote by $\mathcal{L}_Q$ the set of relevant behaviours obtainable via a separable state:
\begin{align} 
p(abc|xyz) = \int \Pi(\lambda) \; p_A (a|x,\lambda ) \; p^{Q}_{BC} (bc|yz,\lambda)  \; d\lambda,
\end{align}
where we have written $p_A (a|x,\lambda )$ rather than $\Tr[A_{a\vert x}\sigma_\lambda^A]$ since any distribution can be obtained with a suitable choice of $\sigma_\lambda^A$ and $A_{a\vert x}$. Denote by $\mathcal{Q}$ the set of quantum distributions $p^{Q}_{BC} (bc|yz,\lambda)$. Unlike $\mathcal{L}_{NS}$, the set $\mathcal{Q}$ has an infinite number of extremal points, and the same approach as above is not possible. We may however do the following. Similarly to \eqref{malin}, replace $p_A (a|x,\lambda )$ via \eqref{replace_det} to obtain 
\begin{align}
p(abc|xyz) & = \sum_i D_A (a|x,i ) [\int \Pi(\lambda)q(i\vert \lambda) \; p^{Q}_{BC} (bc|yz,\lambda) d\lambda] \nonumber \\
&= \sum_i D_A (a|x,i ) \,\tilde{p}^{Q}_{BC} (bc|yz,i),
\end{align}
where since $\mathcal{Q}$ is convex, $\tilde{p}^{Q}_{BC} (bc|yz,i)$ is a sub-normalised distribution of $\mathcal{Q}$, with normalisation $\int \Pi(\lambda)q(i\vert \lambda)d\lambda$ independent of $y,z$. Denote the set of un-normalised distributions of $\mathcal{Q}$ by $\tilde{\mathcal{Q}}$. A solution to the membership problem is thus given by
\begin{align}\label{not_sdp}
&\text{find } \tilde{p}^{Q}_{BC} (bc|yz,i) \;\;\text{such that} \nonumber  \\[8pt]
&p(abc|xyz)=\sum_i D_A (a|x,i ) \,\tilde{p}^{Q}_{BC} (bc|yz,i) \nonumber \\
&\tilde{p}^{Q}_{BC} (bc|yz,i) \in \tilde{\mathcal{Q}}\;\; \forall i, \quad \sum_{bc} \tilde{p}^{Q}_{BC} (bc|yz,i) = \sum_{bc} \tilde{p}^{Q}_{BC} (bc|y'z',i)\; \forall y,z,y',z',i. 
\end{align}
This optimisation is thought to be hard, since no efficient characterisation of the set $\tilde{\mathcal{Q}}$ is known. However, it is known that the set $\mathcal{Q}$ can be characterised from the outside by a sequence of semi-definite programs, known as the NPA hierarchy \cite{npa}. Formally, one defines a sequence of sets $\mathcal{Q}_k$ such that $\mathcal{Q}_1\supseteq\mathcal{Q}_2\supseteq\cdots\supseteq\mathcal{Q}_{\infty}\supseteq\mathcal{Q}$ where membership in $\mathcal{Q}_k$ for any finite $k$ can be tested via an (increasingly large) semi-definite program. This amounts to checking if a matrix $\Gamma$---which is a linear function of the probabilities---is positive semi-definite, and where the constraint $\Gamma_{00}=1$ defines the normalisation of the distributions in the set. If one replaces this normalisation constraint by a positivity constraint $\Gamma_{00}\geq0$, one obtains an analogous hierarchy for un-normalised distributions, whose sets we denote $\tilde{\mathcal{Q}}_k$. One can thus relax the problem \eqref{not_sdp} by relaxing the condition $\tilde{p}^{Q}_{BC} (bc|yz,i) \in \tilde{\mathcal{Q}}$ to $\tilde{p}^{Q}_{BC} (bc|yz,i) \in \tilde{\mathcal{Q}}_k$ for some $k$:
\begin{align}\label{sdp}
&\text{find } \tilde{p}^{Q}_{BC} (bc|yz,i) \;\;\text{such that} \nonumber  \\[8pt]
&p(abc|xyz)=\sum_i D_A (a|x,i ) \,\tilde{p}^{Q}_{BC} (bc|yz,i) \nonumber \\
&\tilde{p}^{Q}_{BC} (bc|yz,i) \in \tilde{\mathcal{Q}}_k\;\; \forall i, \quad \sum_{bc} \tilde{p}^{Q}_{BC} (bc|yz,i) = \sum_{bc} \tilde{p}^{Q}_{BC} (bc|y'z',i)\; \forall y,z,y',z',i.
\end{align}
This problem now has only linear and positive semi-definite constraints and can be solved via a semi-definite program. Since the problem is a relaxation, if this semi-definite program is found to be infeasible, this implies $p(abc\vert xyz)\notin \mathcal{L}_{Q}$, and so the state is entangled. As before, in the case of infeasibility, the dual optimisation problem (also a semi-definite program), provides a separating hyperplane that can be understood as a DI entanglement witness.

\section{Discussion}\label{sec:conclusion}
We have shown that the broadcasting scenario allows for a strong form of Bell nonlocality and DI entanglement certification for a large class of two-qubit states. It is desirable to know the limits of this method. For example, for which values of $\alpha$ is the isotropic state broadcast nonlocal and is this the entire range of entanglement? Note that in the simplest scenario of figure \ref{fig:DAG} where each party has two inputs and two outcomes, all facet inequalities have been enumerated  \cite{Rafael}. Numerical optimisation of these inequalities suggests they do not lead to activation of the isotropic state. Stronger examples than those presented here thus likely require more more inputs or broadcasted parties. For the case of DI entanglement certification, we have shown promising results that suggest DI entanglement certification may be possible in the entire range of entanglement, and it would be interesting to pursue this further. In both cases, the connection between multipartite Bell inequalities and the broadcasting scenario may be useful, as was the case in section \ref{sec:dientcert} that makes use of the Klyshko-Belinskii Bell inequality. %Note that a straightforward application of the Mermin inequality to section \ref{sec:activation} will fail, since the correlations that maximally violate the inequality can be simulated by a non-signalling resources that are product with respect to any bipartition \cite{liang2014anonymous}. 
Another line of investigation could involve generalising the broadcasting scenario so that the transformation device has an input and output, which we have not investigated in this work.  
\smb
One could also consider other resources than non-signalling resources in the definition of broadcast-nonlocality (equation \eqref{modelNS}), as is done in the study of multipartite non-locality \cite{MNLdefs, curchod2015quantifying}. One possibility is to allow the transformation device to prepare signalling resources as well as non-signalling resources, although this is known to cause problems in multipartite scenarios \cite{MNLdefs}. Note that this effectively means the local parties can be considered as a single party (since they can communicate) and so this renders the broadcasting scenario equivalent to the standard Bell scenario in which the transformation and local parties' measurements is equivalent to performing a POVM measurement on the original bipartite state. A more interesting possibility would be to consider the one-way signalling resources (see \cite{curchod2015quantifying,Chaves2017}), which avoid the logical problems that arise with general signalling resources. In this case, we have not established if our examples of activation still hold. We note that for practical applications however, one should not allow for signalling correlations. For example, if one is interested whether or not the state can be replaced by a source of physical classical randomness (that is to say, one in which the $\lambda$'s can be known by an experimenter, such as a commercial random number generator) then any other physical system produced that is conditioned on $\lambda$ must be non-signalling, since otherwise it would allow the experimenter prepare signalling resources on demand. Thus, it only makes sense to consider signalling resources if one assumes that the $\lambda$'s are truly hidden variables, so that they cannot be known even in principle. This is the case for example in the de Broglie–Bohm theory.
\smb
The examples of Bell nonlocality activation of the isotropic state in section \ref{sec:activation} are relative to projective measurements, since for $\alpha\leq0.683$ the isotropic state is known to have an LHV model for projective measurements. Since it is unknown whether a model for general (POVM) measurements exists in the same range, a natural question is whether an example of `genuine' activation with respect to POVM measurements can be achieved in the broadcasting scenario. This is in fact possible by leveraging a previously known result of genuine hidden nonlocality \cite{FlaHNL}, in which it is shown that there exists a family of two qubit states that have an LHV model for general measurements in the standard scenario, and which exhibit hidden nonlocality. Following the connection between hidden nonlocality and the single-copy sequential scenario discussed in section \ref{sec:activation}, one may do the following. Consider a sequential scenario where Bob performs a sequence of two measurements. This can be realised in the broadcasting scenario of figure \ref{fig:DAG} by performing the first measurement at the transformation device, but sending the outcome to Bob, who then outputs this value (he has no input). The post-measurement state is then sent to Charlie. This process is conceptually equivalent to a sequential measurement protocol in which the first measurement device performs a single measurement, where Bob is simply used to output the value of this measurement. Consequently, results in the hidden nonlocality scenario apply to the broadcasting scenario. Thus, the example of hidden nonlocality relative to POVM measurements also implies an example in the broadcasting scenario. Nevertheless, it would be interesting to find stronger examples that rely naturally on the structure of the broadcasting scenario.  
\smb
Our results also suggest the possibility of strengthening existing DI protocols. In particular, since we have shown that DI entanglement certification of the isotropic state at significantly lower visibilities that were previously known, one may hope that this could lead to higher levels of noise tolerance of more complex protocols, such as DI quantum key distribution, DI randomness extraction or self-testing. This warrants additional investigation since---although the state can tolerate more noise---the broadcasting protocol involves additional transformation and measurement devices which themselves will compound further noise to the experimental statistics, and the isometry transformations we find may be difficult to implement in practice. Moving beyond simple entanglement certification, it may also be possible to leverage the semi-definite hierarchy \cite{Moroder2013} to certify not only entanglement, but quantify this via a lower bound on the state negativity.

\section{Conclusion}
In this work we have shown that there exists a simple scenario that allows one to reveal the non-classicality of a large class of two-qubit mixed states. The non-classicality we study takes the form of (i) a type of Bell nonlocality relative to no-signalling hidden variable models, that we call broadcast nonlocality; and (ii) DI entanglement certification. In both cases we show that the broadcasting scenario allows for examples of non-classicality that surpass what is known for all previously studied scenarios that process a single copy of the state. Our work raises many questions that we aim to answer in a future publication, and suggests further applications to DI protocols. 

\section{Acknowledgements}
We would like to thank Marco T\'{u}lio Quintino,  Pei-Sheng Lin, Cristian Boghiu, Marcus Huber, Yeong-Cherng Liang, Gláucia Murta, Charles Xu and Mateus Ara{\'u}jo for valuable comments and discussions.  FH acknowledges funding from the Swiss National Fund (SNF) through the Early Postdoc Mobility fellowship P2GEP2$\_$181509. JB and DC acknowledges funding from the Spanish MINECO (Severo Ochoa SEV-2015-0522), Fundacio Cellex and Mir-Puig, Generalitat de Catalunya (SGR 1381 and CERCA Programme). JB acknowledges funding from the AXA Chair in Quantum Information Science. DC acknowledges the Ramon y Cajal fellowship.

\bibliography{bibliography.bib}

\newpage

\begin{appendices}

\section{Linear optimisation of visibilities} \label{linpr}

Here we consider calculating the robustness to noise for behaviours generated via a family of states with a linear parameter, with fixed local measurements and channel(s). For example, consider the family of states
\begin{align}\label{gen_state}
    \rho_v = v \rho_{ent} + (1-v) \rho_{noise}
\end{align}{}
with $0 \leq v \leq 1$, and where $\rho_{ent}$ is an entangled bipartite state, while $\rho_{noise}$ is a separable state (for example, the isotropic state \eqref{isotropic2} is already in that form). Fix a channel $\Omega_{B_0\rightarrow BC}$ and local measurements $A_{a\vert x}, B_{b\vert y}, C_{c\vert z}$. The behaviour is thus given by

 \begin{align}
   p_v(abc|xyz) &=  \Tr (  A_{a|x} \otimes B_{b|y} \otimes C_{c|z} \; \mathbb{1} \otimes \Omega_{B_0\rightarrow BC} (\rho_v) )  \nonumber\\
   &=v\;p_{ent}(abc|xyz)+(1-v)p_{noise}(abc|xyz), \label{linear_prob}
\end{align}{}
where 
\begin{align}
    p_{ent}(abc|xyz)=\Tr(A_{a|x} \otimes B_{b|y} \otimes C_{c|z} \; \mathbb{1} \otimes \Omega_{B_0\rightarrow BC} (\rho_{ent}))
\end{align}
and similarly for $p_{noise}$. Since $p_v(abc|xyz)$ is linear with respect to $v$, one can find the exact value $v^*$ for which the behaviour becomes nonlocal using the following linear program (using \eqref{LP}):
\begin{align}\label{linprog} &\text{maximise} \; v \quad \text{such that} \\
    \nonumber   &v\;p_{ent}(abc|xyz)+(1-v)p_{noise}(abc|xyz)    =  \sum_{i,j} p_{ij} \; D_A(a\vert x ,i) \; E_{BC}(bc|yz,j) \;\;\forall a,b,c,x,y,z. \\
    &\nonumber p_{i,j} \geq 0 
\end{align}
The optimal dual variables to the euqality constraints provide a Bell-type inequality that is violated by $p_v(abc|xyz)$ for $v>v^*$.

\section{Heuristic optimisation of visibility}\label{ap:heuristic} 

Here we detail the algorithm used for the heuristic minimisation of state visibility. Consider again a one-parameter family of state \eqref{gen_state}. We want to find the lowest visibility $v$ such that the state is nonlocal (in some scenario), that is, the general task is

\begin{align} &\text{minimise} \; v \\
    \nonumber &\text{s.t.}  \; \rho_v = v \rho_{ent} + (1-v) \rho_{noise} \; \text{is nonlocal} .
\end{align}{}
In order to show that the state is nonlocal in the standard Bell scenario one needs to violate a Bell inequality, i.e., one needs measurements $\{A_{a|x} \}$ and $\{B_{b|y}\}$ such that behaviour $p(ab|xy) = \Tr ( A_{a|x} \otimes B_{b|y} \; \rho )$ violates a Bell inequality. In the broadcasting scenario of e.g.\ figure \ref{fig:DAG} one needs to find a map $\Omega_{B_0\rightarrow BC}$, and measurements $\{A_{a|x} \}$, $\{B_{b|y}\}$ and $\{C_{c|z}\}$ such that behaviour

 \begin{align}
   p(abc|xyz) =  \Tr (  A_{a|x} \otimes B_{b|y} \otimes C_{c|z} \; \mathbb{1} \otimes \Omega_{B_0\rightarrow BC} (\rho_v) )  
\end{align}
 violates an inequality tailored to the broadcasting scenario (for example inequality \eqref{ineq_1Alice}). Below, we show how to optimise the quantum value of such inequalities.
 
\smb
 
\subsection{Seesaw procedure for maximising inequality violation} 
 
Let us fix an inequality in the broadcasting scenario, for instance with three parties as in figure \ref{fig:DAG} (where we have renamed the input/output labels):

\begin{align}
   \sum_{a_1 a_2 a_3 x_1 x_2 x_3} \mathcal{I}_{a_1 a_2 a_3}^{x_1 x_2 x_3}\; p(a_1 a_2 a_3|x_1 x_2 x_3) \leq L
\end{align}{}
where the local bound $L$ depends on the underlying model one wants to rule out (i.e.\ the scenarios of sections \ref{sec:activation} or \ref{sec:dientcert}). For a fixed strategy, the quantum value is given by 

 \begin{align}
      \sum_{a_1 a_2 a_3 x_1 x_2 x_3} \mathcal{I}_{a_1 a_2 a_3}^{x_1 x_2 x_3}  \;\Tr (  M^{(1)}_{a_1|x_1} \otimes M^{(2)}_{a_2|x_2} \otimes M^{(3)}_{a_3|x_3} \; \mathbb{1} \otimes \Omega_B (\rho_{AB}) )  
\end{align}{}
Below we propose an algorithm to optimise the quantum violation over local measurements and map(s), for a fixed inequality and state $\rho_{AB}$.

\mb

\noindent\textbf{Seesaw for optimising inequality violation for fixed state} \label{seesaw}

\smb 

Assume one want to maximise the violation of a general inequality $\mathcal{I}_{a_1 a_2 \cdots a_n}^{x_1 x_2 \cdots x_n}$ (in a general broadcasting scenario) for a fixed state $\rho_{AB}$ and (for now) fixed channels $\Omega_{A/B}$, and sets of measurements $M^{(1)}_{a_1|x_1}$, $M^{(2)}_{a_2|x_2}$, ... $M^{(n)}_{a_n|x_n}$. One can do the following. 

\begin{enumerate}

    \item  Fix randomly all sets of measurements 
    \item  Set $k$=1
    \item Optimise the inequality with respect to set of measurements $M^{(k)}_{a_k|x_k}$, update variables $M^{(k)}_{a_k|x_k}$ accordingly
    \item  Repeat point 3 for $k = 2 ... n$
    \item  Repeat point 2 - 4 until two successive values of the inequality are equal up to some small precision $\epsilon$.
\end{enumerate}

The key point is to notice that point 3 corresponds to optimising a linear function, with SDP constraints (namely $M_{a_k|x_k}\geq 0$ and $\sum_a M_{a_k|x_k} = \mathbb{1} \, \forall x_k$), that is, it can be framed as a semi-definite program, for which there exist efficient algorithms \cite{Vandenberghe1996}. 
\smb
One also needs to optimise over the channel(s). This can be done via the Choi-Jamiolkowski isomorphism \cite{Choi}. Taking the case of figure \ref{fig:DAG} of section \ref{sec:activation} as a concrete example, for the map $ \Omega_{B_0\rightarrow BC}$ one has
\begin{align}
     \Omega_{B_0\rightarrow BC} (\sigma_{B_0}) = \Tr_1 (\rho_{\Omega} (\sigma_{B_0}^{T} \otimes  \mathbb{1}_{BC}) )
\end{align}
where $\sigma_{B_0}$ is some operator acting on $\mathcal{H}_{B_0}$, and $\ket{\Phi^+}$ is the maximally entangled state of local dimension $d=\text{dim}(\mathcal{H}_{B_0})$ so that $\rho_{\Omega_{B_0\rightarrow BC}}=d\cdot\mathbb{1} \otimes \Omega [\proj{\Phi^+}]$ is the Choi state of map $\Omega_{B_0\rightarrow BC}$. The Choi state thus satisfies $\rho_{\Omega} \geq 0$ and $\Tr_{BC} ( \rho_{\Omega} ) = \mathbb{1}_{B_0} $. That is, variable $\rho_{\Omega}$ can be treated as an SDP variable. Furthermore, the reverse is true: any operator satisfying the SDP conditions is the Choi state of some channel. The probabilities can thus be written
\begin{align}
    p(abc|xyz)=\Tr (  (A_{a|x}\otimes \mathbb{1}_{B_0} \otimes B_{b|y} \otimes C_{c|z}) \; (\mathbb{1}_A \otimes \rho_{\Omega})(\rho_{AB_0}\otimes \mathbb{1}_{BC}))
\end{align}
This expression is linear in the Choi state, which means for fixed state and measurements, the inequality can be maximised over all quantum channels of the given dimension. A similar procedure allows the same to be done for Alice's channel (if she has one). Combining this with the above method to optimise the measurements, one can `see-saw' between the two until convergence has been reached, which in practice performs well for simple scenarios. 

\subsection{Global heuristic procedure}

Relying on the previous section and on a seesaw procedure described in Appendix \ref{seesaw}, we describe here a global heuristic procedure to minimise the critical visibility of a fixed family of states. 

\begin{enumerate}

    \item  Set random measurements $\{M_{a_i|x_i}\}$ and channels $\Omega_{A}$, $\Omega_B$
    \item  Compute $v^*$ (using the linear program \eqref{linprog})
    \item Repeat step 1-2 until $v^* < 1$, and keep the corresponding inequality $\mathcal{I}$
    \item Set k=0
    \item  Optimise the violation of $\mathcal{I}_0$ for state $\rho_{v^*}$ (using seesaw from \ref{seesaw} )
    \item  Compute new $v^*$ using optimal measurements and map found in 5 (using linear program \eqref{linprog}), and keep corresponding inequality $I_{(k+1)}$
    \item Set $k=k+1$
    \item Repeat steps 5-7 until two successive $v^*$ are equal up to some small precision. 
\end{enumerate}

Using this procedure in the symmetric case described in \ref{sec:symmetric4}, with the two-qubit isotropic state and using $3$ inputs for parties A and C, $2$ for B and D, and $2$ outputs for each, we found a violation for $\alpha \simeq 0.559$, showing that activation for values of $\alpha$ lower than $1/\sqrt{3}$ is indeed possible in the broadcasting scenario.   

In this example, the two local channels applied to the state are identical and are again given by \eqref{isometry}, while the local measurements as well as the optimal inequality are given as auxiliary files. We suspect that even stronger examples could be found with further investigation, and it is still unknown whether activation in the entire range of entanglement is possible. 

\end{appendices}

\end{document}